# Objectives of platform research: A co-citation and systematic literature review analysis

Fabian Schüler*

IPRI - International Performance Research Institute gGmbH,

Königstr. 5, 70173 Stuttgart, Germany

E-Mail: fschueler@ipri-institute.com

Petrik Dimitri

Graduate School of Excellence advanced Manufacturing Engineering,

Nobelstr. 12, 70569, Stuttgart, Germany

E-Mail: dimitri.petrik@gsame.uni-stuttgart.de

* Corresponding author

**Abstract**

Business economics research on digital platforms often overlooks existing knowledge from other fields of research leading to conceptual ambiguity and inconsistent findings. To reduce these restrictions and foster the utilization of the extensive body of literature, we apply a mixed methods design to summarize the key findings of scientific platform research. By combining a co-citation bibliometric analysis with a systematic qualitative content analysis, our first result presents an overview of 14 platform-related research fields, obtained from 10.357 research papers from the Web of Science database. Building upon this cluster analysis, we employ qualitative content analysis techniques to the platform ecosystem research field, which contains 107 papers. The second result aims to identify influential research objectives and explores how they evolved. This consequently leads to a comprehension of different platform conceptualizations and extends the quantitative insights of the study. During the content analysis, we identify three primary research objectives related to platform ecosystems: (1) general literature defining and unifying research on platforms; (2) exploitation of platform and ecosystem strategies; (3) improvement of platforms and ecosystems. By applying the business economics research lens, this paper develops a discussion on the identified insights and promising future research directions to enhance business economics and management research on digital platforms and platform ecosystems. Our study employs a mixed methods bibliometric study to provide a holistic knowledge structure of the existing platform research, helping the business economics research community and other disciplines to avoid ambiguity and offering an overall picture of platform research.

**JEL-Classification: L15,M15,O32**

**Zusammenfassung:**

Die betriebswirtschaftliche Forschung zu digitalen Plattformen vernachlässigt häufig vorhandenes Wissen aus anderen Forschungsbereichen, was zu konzeptioneller Ambiguität und inkonsistenten Ergebnissen führt. Um diese Einschränkungen zu verringern und die Nutzung der umfangreichen bestehenden Literatur zu fördern, wenden wir einen Mixed Methods Ansatz an, um die wichtigsten Ergebnisse der wissenschaftlichen Plattformforschung zusammenzufassen. Durch die Kombination einer bibliometrischen Co-Zitationsanalyse mit einer systematischen qualitativen Inhaltsanalyse bietet unser erstes Ergebnis einen Überblick über 14 plattformbezogene Forschungsfelder, die aus 10.357 Forschungsarbeiten aus der Web of Science-Datenbank identifiziert wurden. Aufbauend auf dieser Clusteranalyse setzen wir qualitative Inhaltsanalysetechniken für das Forschungsfeld der Plattform-Ökosysteme ein, das 107 Artikel umfasst. Dieses zweite Ergebnis zielt darauf ab, die relevanten Forschungsobjekte in diesem Forschungsstrang zu identifizieren und zu untersuchen, wie sich diese im Laufe der Zeit entwickelt haben, um ein vollständiges Verständnis der Konzeptualisierung der digitalen Plattformen zu ermöglichen und die quantitativen Erkenntnisse zu erweitern. In unserer Inhaltsanalyse identifizieren wir drei grundlegende Forschungsziele in Bezug auf Plattform-Ökosysteme: (1) allgemeine Literatur, die Plattformen definiert und die verschiedenen Forschungsansätze vereinheitlicht; (2) die Nutzung von Plattform- und Ökosystemstrategien; (3) die Verbesserung von Plattformen und deren Ökosystemen. Aufbauend auf einer betriebswirtschaftlichen Perspektive wird in diesem Artikel eine Diskussion über die identifizierten Erkenntnisse und vielversprechende zukünftige Forschungsrichtungen zur Verbesserung der betriebswirtschaftlichen Plattformforschung entwickelt. Unsere Studie nutzt einen bibliometrischen Mixed Methods Ansatz, um eine ganzheitliche Wissensstruktur der vorhandenen Plattformforschung zu erarbeiten, die Forschern der Wirtschaftswissenschaften und anderen Disziplinen hilft, durch ein Gesamtbild der Plattformforschung Ambiguität zu vermeiden.

*Schlüsselwörter: digitale Plattform, Plattformökosysteme, Forschungsschwerpunkte, Literaturanalyse, Co-Zitationsanalyse*

# 1. Research on digital platforms

In recent years, the term platform has become nearly ubiquitous and emerged in various fields like product development, operations management, technology strategy, and industrial economics (Gawer, 2014). Digital platforms exist in various forms and domains, ranging from social networks to cross-company Internet of Things (IoT) platforms. Despite its massive significance, there is no universal definition for the term digital platform (Reuver, Sørensen, & Basole, 2018). Various scientific communities and disciplines developed distinctive perspectives on digital platforms (see section 3).

The economic fundamentals developed by Rochet and Tirole (2003) show that markets on which network effects operate have multiple sides, which in turn are connected by a platform. In this context, the platform acts as an intermediary and enables interactions (Rochet & Tirole, 2006, p. 647). On a general level, most researchers on platforms agree on understanding platforms as an intermediary organizational form between two or more sides, providing the necessary infrastructure to enable interactions between different user groups (Armstrong, 2006; Cusumano & Gawer, 2002; Gawer & Cusumano, 2014; Rochet & Tirole, 2003). In their management book, Gawer and Cusumano (2002) explain how companies can use digital platforms for competitive advantages and define leverage mechanisms for companies to become platform leaders. Gawer and Henderson (2007, p. 6) define a platform "as a firm that owns a core element of the technological system that defines its forward evolution". In this context, platforms are seen as a modular technological infrastructure to enable digital services (e.g. through complementary applications) and create additional value for the actors, which are using the platform to innovate (Baldwin & Woodard, 2009; Hein et al., 2019). This paradigm change requires platform companies to implement complex organizational, cultural and technological changes to externalize specific innovation capabilities to independent actors, while focusing internal resources and capabilities, towards the development of the platform and the aligning ecosystem (Gawer & Cusumano, 2014; Hein et al., 2019; Yoo, Henfridsson, & Lyytinen, 2010). Despite a period of nearly 18 years since the platform concept was introduced, a comprehensive overview of the mechanisms how to maintain attractive platforms is still lacking. While some digital platforms have tremendous success, the failure rate is also surpassingly high (Van Alstyne, Parker, & Choudary, 2016). Besides, the absence of clear market leaders in certain domains, such as IoT (Krause et al., 2017), indicates how challenging it is to establish a platform.

Not surprisingly, the broad research focus on digital platforms has led to literature reviews in various fields of research. McIntyre and Srinivasan (2017) have analyzed the body of literature on digital platforms from different perspectives and match the current findings to industrial organization economics, technology management, and a strategic management perspective. Other reviews focus on the processes and structures of multisided platforms and their impact on technology (Tan, Pan, Lu, & Huang, 2015). Reuver et al. (2018) analyze research connections between digital platforms and information systems research and identify various research questions among these topics. Regarding our research focus, we also found papers that aim to combine different research perspectives into a more holistic view of digital platforms (Gawer, 2014).

*Table 1: Existing high impact literature reviews associated with digital platforms[1]*

| Authors | Research focus | Purpose and findings |
|---|---|---|
| Gawer and Cusumano (2014) | Strategic Management perspective on industry platforms | Literature review on industry platforms and their effect on managing innovation within and outside the firm. |

---

[1] According to VHB-JOURQUAL 3 categories A+, A and B; see https://vhbonline.org/vhb4you/vhb-jourqual/vhb-jourqual-3/gesamtliste

| Thomas, Autio, and Gann (2014) | Strategic Management perspective on platforms | Literature review identifying four distinct research streams associated with platforms: organizational platforms, product family platforms, market intermediary platforms, and platform ecosystems. |
|---|---|---|
| Tan et al. (2015) | Information Systems perspective on platforms | Literature review analyzing the role of IS capabilities in the development of multi-sided platforms. |
| McIntyre and Srinivasan (2017) | Strategic Management perspective on platforms | Literature review analyzing platform research from the industrial organization economics, technology management, and strategic management perspective. |
| Reuver et al. (2018) | Information Systems perspective on platforms | Literature review developing a research agenda for digital platform research in information systems. |

Recent platform research mentions the lack of general theories and empirical data concerning platforms and aligning ecosystems (Jacobides, Cennamo, & Gawer, 2018; Reuver et al., 2018). As platforms are increasingly getting more connected and expand into larger digital ecosystems, digital platforms are becoming even more complex research objects. To improve understanding of these complex ecosystems, this study analyzes the differing understandings and perspectives of digital platforms, uncovering previously conducted research efforts and remaining gaps for future research.

In order to create a more systematic and encompassing picture of the platform research agenda, especially building on the platform concept of Gawer (Gawer, 2014; Gawer & Henderson, 2007), we engage in a systematic quantitative review of the existing literature on platforms. Applying a mixed methods design and using bibliometrics in combination with qualitative content analysis, our primary goal is to (i) identify the theoretical foundations of platform research, (ii) structure subsequent research, and (iii) reveal open research gaps. Our bibliometric approach is based on co-citation analysis to reveal essential papers, which are connected to Gawer's research perspective on platforms. Furthermore, to the best of our knowledge, no bibliometric studies on digital platforms, as defined by Gawer and Cusumano (2002), exist until January 2020, aiming to describe central existing research lenses. The study conducted by Suominen, Seppänen, and Dedehayir (2019) differs from ours in terms of the research subject (innovation ecosystems). The same applies to the study of Fu, Wang, and Zhao (2018), which analyzes a similar body of literature but focuses on platform-based service innovation. Given the heterogeneity of the platform research streams and the absence of a holistic examination, we sense an opportunity to enhance the understanding between the different research perspectives. Providing an overview of essential papers, central concepts, and mechanisms of platforms, which are represented by distinctive research lenses, our results support researchers in economic research disciplines to complement recent research on digital platforms mainly developed in areas of information systems and computer science.

The remainder of this paper is subdivided into five parts. In the following section, we introduce our methodology and describe how we gathered the data. In Section 3, we present the derived co-citation clusters, describe the network, and provide a detailed analysis of the second cluster, which fits our platform perspective. Section 4 discusses the implications of our derived clusters and develops future research directions. Lastly, we close with our conclusion and limitations of the study in Section 5.

## 2. Method and data

### 2.1. Method

In this study, we combine bibliometric methods and inductive coding of the literature in a specific cluster to gain comprehensive insights into existing research on platforms. Mixed methods integrate qualitative and quantitative research methods in a sequential way to benefit from a holistic understanding of a phenomenon (Venkatesh, Brown, & Bala, 2013). The combination of the methodologies is used to compensate for the respective weaknesses of the methods (Raghuram et al., 2010; Wang et al., 2016). Hence, this paper applies a mixed methods design to a quantitative bibliometric study, extending it by a follow-up qualitative analysis.

Bibliometric analysis uses statistical methods to discover publication patterns (McBurney & Novak, 2002) and builds comprehensive pictures of research landscapes, identifies specific topics within these landscapes, and explores their evolutionary development. Application-oriented research of this method shows its application in various contexts in business economics research. Moreover, bibliometric methods use specific software tools and defined algorithms (see section 3) that bring the advantage of objectivity (Raghuram, Tuertscher, & Garud, 2010; Wang et al., 2016). Structured literature reviews are affected by time constraints and the subjectivity of the researcher, usually presenting only a particular perspective on the researched topic (Chen, 2016). The software support, the objectivity, and the quantifiability of bibliometric methods reduce the effect of subjectivity, delivering a more balanced picture on digital platforms (Chen, 2016), which is complemented by the subjective interpretation of the studied literature (Wang et al., 2016). Therefore, this study complements previously published qualitative literature reviews on digital platforms.

We use a bibliometric approach and a computerized gathering of data to map and analyze a large number of publications referring to platforms. Pritchard (1969) shaped the term bibliometrics for the application of statistical and mathematical methods to all means of communication. Therefore, bibliometrics is a quantitative analysis of literature to study the connections and patterns between different articles based on their citations. A citation occurs, when one paper refers to another. Bibliometrics can assess the scientific accomplishments of researchers or institutions and their impact in an explorative way, to gain new insights into existing fields of research (Zupic & Čater, 2015). Citation analysis is valuable to identify source papers, influential papers, as well as the relationships and development paths among related papers. This method has been extensively used to address similar research questions in many disciplines of social sciences and is the most used technique to map research areas (Zhao & Strotmann, 2015), but has not yet been used for a holistic mapping of prior digital platform research.

For our study, we decided to use co-citation analysis instead of bibliographic coupling because it explains which publications are connected on purpose by the authors. Although bibliographic coupling serves the goal to discover to what extent the papers share a common background, the usefulness of this methodology is controversial, because it does not consider the changing coupling strength over time. We use co-citation analysis because its design improves several shortcomings of bibliographic coupling and represents semantic similarity. It has been successfully validated (Small, 1973) and was competitive against other measures (Boyack & Klavans, 2010). Finally, a co-word analysis of the abstracts is used to label the research clusters, which emerged during the analysis.

Small (1973) introduced co-citation analysis as a method for the evaluation of semantic similarities of papers that share citations. A co-citation is defined as an occurrence in which two papers are cited together in another paper. The more co-citations two papers receive, the higher their co-citation strength, and the more likely they are semantically related (Small, 1973). Figure 1 shows three citing papers (A, B, C) and the resulting co-citation values ($Z_{ij}$) for the papers 1, 2, 3, and 4. After the calculation of the co-citation values for all papers, the publications are organized in a network. In this network, every publication is a node. As defined by Small (1973), the edges of the network between two publications $i$ and $j$ are defined by the standardized co-citation strength $w_{ij} = \frac{Z_{ij}}{\sqrt{Z_i Z_j}}$.

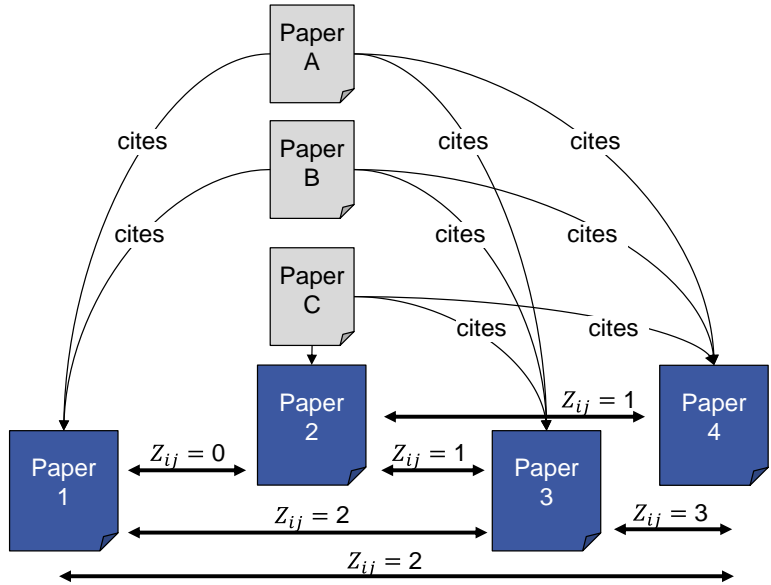

Figure 1: Simplified example of the co-citation value $Z_{ij}$

As visualized in Figure 1, the co-citation between two publications counts the number of other papers that cite both publications together. The general idea is that the meanings are closely related when two papers are often cited together. For the construction of the network, all edges between the publications are calculated. In the final network, publications that have a strong connection (high co-citation value) are close to each other, whereas the distance between publications with a slight connection is maximized. Figure 2 explains this coherence, papers 3 and 4 are close and have a high co-citation value while the distance between papers 1 and 2 is maximized, due to their co-citation value of zero.

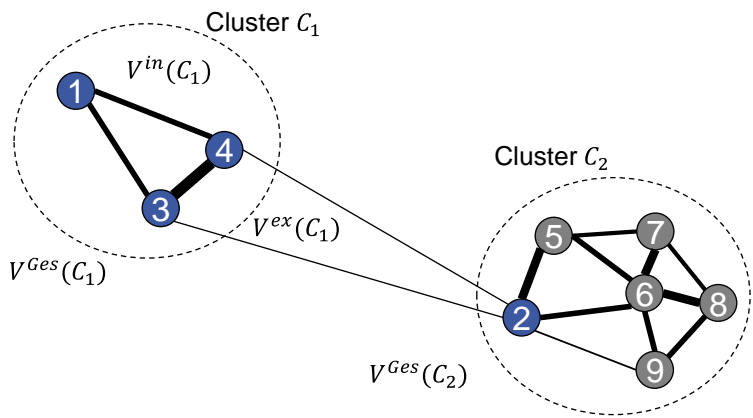

Figure 2: Formation of clusters and definition of cluster equations.

For the identification of semantically similar publications, the network is separated into clusters. In manually defined research fields, the semantic relations between publications are strong compared to the relations to publications in other fields. The equivalent is represented in a network when the connectedness between nodes inside a cluster is high while the connectedness between clusters is low. Let $v^{in}(C)$ be the connectedness of the node pairs in a cluster and $v^{ex}(C)$ be the connectedness between the clusters. Similarly, the sum $v^{in}(C)$ and $v^{ex}(C)$ is represented by the total connectedness $v^{tot}(C)$. This valuation is executed by spectral clustering because it is a network-based cluster algorithm fulfilling the specified requirements. The algorithm iteratively maximizes the total cluster quality $Q$ and minimizes cluster openness $E$ of all clusters. Table 2 explains and defines $Q$ and $E$. The relationships between the values for connectedness and transformation of the connectedness variables in q and e is presented in Table 2:

Table 2: Equations for connectedness values, cluster quality, and cluster openness

| Total connectedness | $v^{tot}(C) = \sum_{i \in C} \sum_{j \in D} w_{ij}$ | Maximized by the algorithm | $Q = \sum_{k=1}^{K} V^{in}(C_k)$ |
|---|---|---|---|
| External connectedness | $v^{ex}(C) = \sum_{i \in C} \sum_{j \in D/C} w_{ij}$ | Minimized by the algorithm | $E = \sum_{k=1}^{K} \frac{V^{ex}(C_k)}{V^{tot}(C_k)}$ |
| Internal connectedness | $v^{in}(C) = \sum_{i \in C} \sum_{j \in C} w_{ij}$ | | |

For further details on the algorithm, see Chen, Ibekwe-SanJuan, and Hou (2010). Because two values are simultaneously optimized, the solution is not necessarily unique. The number of clusters must be predefined for the algorithm. For the identification of the optimal number of clusters, clustering is performed for different numbers of clusters, and the algorithm selects the optimal number. For a quantitative understanding of the results, text mining (Log-Likelihood Ratio) is used to name the clusters. We used the software CiteSpace Version 5.6.R2 (Chen & Song, 2019).

After the systematic identification of the relevant literature, we examine the most appropriate research cluster in more detail, applying a qualitative literature review. The literature analysis has the purpose of analyzing the identified literature regarding the development of platform research as a topic of interest. In order to consolidate existing research, we use inductive coding to build sub-clusters and summarize the findings, keeping in mind the delicate balance between the relevance and detailed description of each paper (Fisch & Block, 2018). Besides, this step allows the identification and separation of papers which are not directly linked to platform research (see sub-cluster 4). Afterward, we describe the platform-related concepts in a structured way. We also structure the variety of research topics in a concept matrix as proposed by Webster and Watson (2002) and provide additional textual explanations for each concept. The following figure 3 illustrates the process of our analysis:

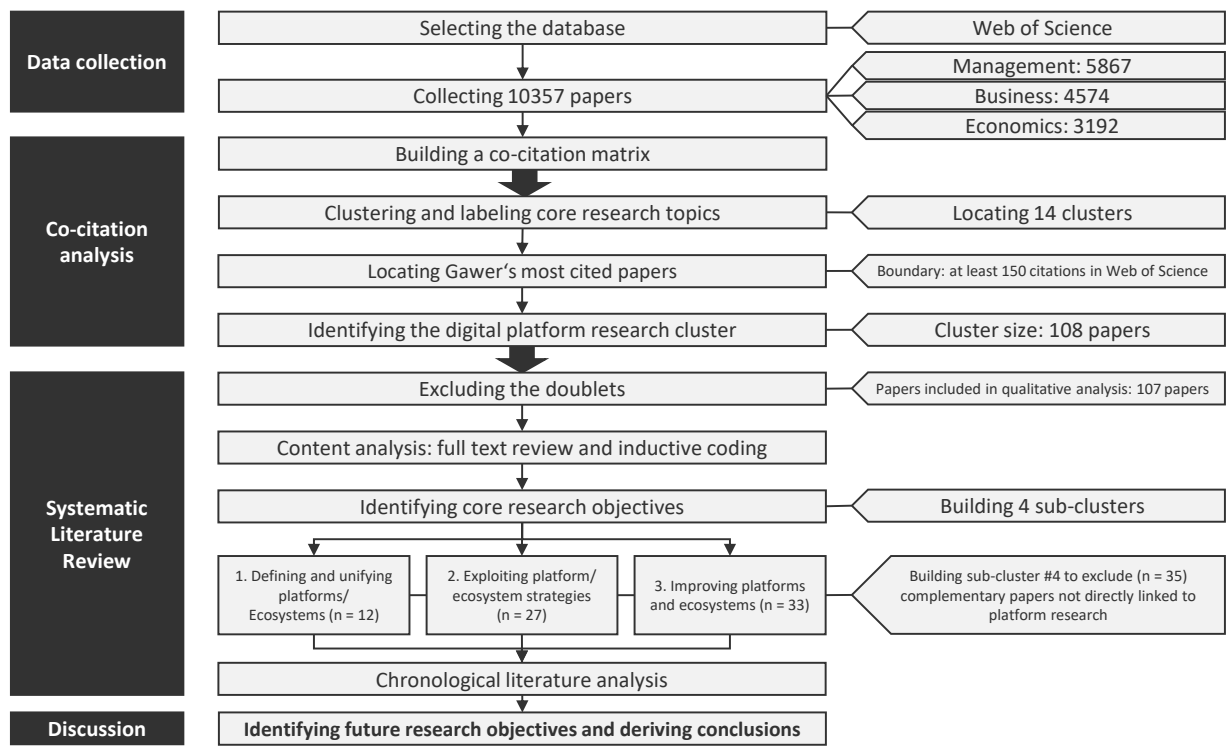

*Figure 3: Overview of the mixed methods process and the resulting body of literature*

## 2.2. Data

The data set of the bibliometric analysis is a collection of publications that refer to platforms. Each data point represents a publication by containing the full reference of the publication, including title, author, year, and relevant information for citing the journal article. The data was collected from the Web of Science database (14.01.2020). We used Web of Science because it is one of the largest scientific publication databases (Falagas, 2008). Furthermore, Web of Science and its most significant competitor Scopus deliver highly correlated results (Archambault, Campbell, Gingras, & Larivière, 2009). We searched for the term "platform" in the search fields, title, abstract, and keywords. Because the search intends to explore new research directions, we included as many publications as possible. Although the search targeted mainly publications of digital platforms, we did not restrict the search term to "digital platform", because many authors omit the usage of the adjective "digital". Despite the superfluous articles that use the platform term synonymously (e.g. sharing economy or crowdfunding), no restrictions, based upon additional keywords, were imposed in order to obtain a complete picture of past platform research. However, we define the review period between 2002 and 2019, as our perspective of platform research has emerged and developed during this period (Cusumano & Gawer, 2002).

Based on the settings explained above, the search collected a data set of 10.357 different publications for the bibliometric analysis. The selected categories are *Management* (5.867 publications), *Business* (4.574 publications), *Economics* (3.192 publications). The whole data set is also available here: https://bit.ly/38JcELA.

# 3. Results

## 3.1. Co-citation analysis

Overall, using the co-citation approach, we identify 14 research areas (e.g. clusters). Figure 4 presents all clusters which have emerged in the analysis. As described above, we apply the platform concept of Gawer for this analysis and found that cluster #2 "platform ecosystem" encompasses all scientific publications by Anabelle Gawer which have more than 150 citations in the Web of Science Core Collection (Gawer, 2014; Gawer & Cusumano, 2008, 2014; Gawer & Henderson, 2007) as well as her famous management book "Platforms, markets and innovation" (Gawer, 2009). As this cluster addresses our research focus, we proceed with a detailed analysis of the 107 papers and books included in cluster #2 "platform ecosystem". The average publication year for cluster #2 is 2011. A full list of all papers from cluster #2 is available here: https://bit.ly/2S0CmnY.

Analyzing the resulting network of our analysis, cluster #2, shows deep connections to the "two-sided market" cluster (see Figure 4). This cluster covers the economic foundations of platform research and is mentioned in most publications in the platform ecosystem cluster, as assumed by McIntyre and Srinivasan (2017) or Gawer (2014). Cluster #5 "manufacturing processes", with an average publication year of 1998 is much older than cluster #2 (2011) and provides fundamental insights about product modularity and the subliming architecture, which have a high impact on the platform perspective developed by Gawer. Cluster #3 provides research about open innovation and collaboration with a strong link to open source software (Hippel, 2006; Tapscott & Anthony, 2011). This cluster provides insights into various aspects of how to increase the value-added into an open software ecosystem by different users or development partners. While the upper right part of the map is older than the platform ecosystem cluster (#3=2004, #4=2006, #5=1998, #9=2003, #11=2004) and has influenced research on platform ecosystems, the lower left part has mainly developed after the establishment of cluster #2. Most of these clusters, like social media (#2), sharing economy (#6), equity crowdfunding (#7), crowdsourcing platform (#8), and gig economy (#15), represent specific use cases of platform ecosystems. Future research might analyze the specific mechanisms developed in these clusters on their generalizability for other platform ecosystems. The same applies to the disruptive innovation cluster (#10), which focuses on business models and business model innovation research. This cluster is also younger than cluster #2 and encompasses all sorts of business models, which can be enabled via platforms. This lens may also provide structured approaches to support the development of thriving platform ecosystems for both the platform provider and the complementing companies. A detailed analysis of cluster #1 might also improve future research on platform ecosystems. The electronic word of mouth cluster might offer relevant knowledge about marketing strategies and social media mechanisms for the growth of platform ecosystems. The isolated "user anonymity" cluster (#17) consists of methodical research on multivariate data analysis and structural equation modeling and covers the methodical foundations of the empirical papers. An overview of the ten most essential papers in each cluster is available here: https://bit.ly/36B2o6u.

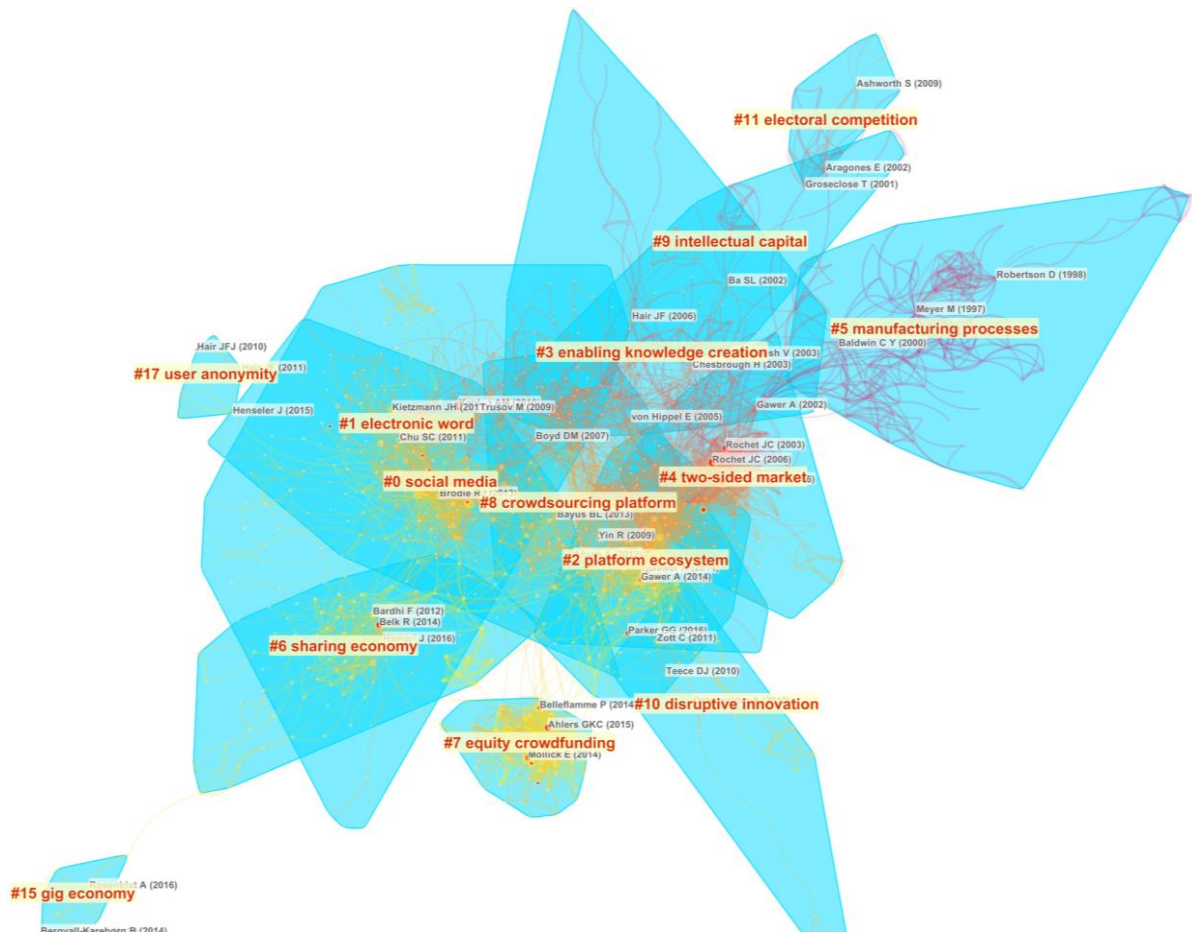

*Figure 4: Research clusters including the term "platform" in management, business, and economic research*

## 3.2. Systematic literature review

After a detailed content-related analysis of all papers contained in the main cluster #2 "platform ecosystem", we divide the cluster into four substantially different sub-clusters (see Table 3). While other papers found conceptual distinctions between different lenses on the research on digital platforms building on different scholars, we divide the papers included in cluster #2 into different research objectives. Sub-cluster 1 includes all publications, which aim to define or unify research on platform ecosystems. Additionally, these papers provide valuable information about the main components of platform ecosystems as well as the underlying mechanisms of value creation from a conceptual perspective. Sub-cluster 2 covers all papers, which analyze how companies can benefit from implementing platform strategies. Sub-cluster 3 includes mostly quantitative and qualitative research and aims to improve existing platforms by analyzing different characteristics and coherences. Sub-cluster 4 comprises studies, which had a high impact on the research on platform ecosystems but has no direct linkage to the issues addressed in the three categories previously described. In the following section, we provide a detailed overview of the research objects and future research opportunities for sub-clusters 1, 2, and 3.

*Table 3: Overview: Sub-clustering of the platform ecosystem cluster for an in-depth analysis*

| Sub-cluster names / Metadata | 1. Defining and unifying platforms/ecosystems | 2. Exploiting platform/ ecosystem strategies | 3. Improving platforms and ecosystems |
|---|---|---|---|
| # of publications | 12 | 27 | 33 |
| # on management books | 1 | 5 | 2 |
| Avg. publication year | 2012 | 2010 | 2012 |
| Most frequently cited (Top 5) | Gawer (2014); Tiwana, Konsynski, and Bush (2010); Thomas et al. (2014); McIntyre and Srinivasan (2017); Baldwin and Woodard (2009) | Gawer and Cusumano (2014); Parker, van Alstyne, and Choudary (2016); Eisenmann, Parker, and van Alstyne (2011); Ceccagnoli, Forman, Huang, and Wu (2012); Yoo, Boland, Lyytinen, and Majchrzak (2012) | Adner and Kapoor (2010); Zhu and Iansiti (2012); Boudreau (2012); Boudreau (2010); Cenamor, Usero, and Fernández (2013); |
| Most recently published (Top 5) | Reuver et al. (2018); Jacobides et al. (2018); Adner (2017); McIntyre and Srinivasan (2017); Gawer (2014) | Evans and Schmalensee (2016); Eloranta and Turunen (2016); Van Alstyne et al. (2016); Brouthers, Geisser, and Rothlauf (2016); Parker et al. (2016) | Zhu and Liu (2018); Kazan, Tan, Lim, Sørensen, and Damsgaard (2018); Parker and van Alstyne (2018); Kapoor and Agarwal (2017); Parker, van Alstyne, and Jiang (2017) |
| Most common journals | Strategic Management Journal; Information Systems Research | Harvard Business Review; MIT Sloan Management Review; Academy of Management Review | Management Science; Strategic Management Journal; Organization Science; MIS Quarterly |
| User/Complementor perspective | 0 | 2 | 6 |
| # of qualitative studies # of quantitative studies # of conceptual papers # of theoretical models | 0 0 11 0 | 4 4 20 4 | 10 16 12 10 |

Sub-cluster 1 contains 11 influential papers focusing on different understandings and definitions of platforms and ecosystems. The primary target of the publications in this sub-cluster is to shape a better understanding of the multifaceted phenomenon of platform ecosystems. This attitude is mainly embodied by Gawer (2009) and her Book "Platforms, Markets and Innovation", which brings together different approaches that foster a better understanding of platform ecosystems. This book is the oldest publication in this sub-cluster and provides a foundation for a general understanding of platform ecosystems by illustrating all major perspectives.

Building on prior research, Gawer's fundamental organizational framework on technological platforms provides a comprehensive understanding of platforms, identifying three distinctive forms of platform ecosystems (Gawer, 2014). Industry platforms form the most open class of technology platforms: providing open interfaces platforms are building blocks that enable independent actors to develop complementary products and services in a platform-based ecosystem. The interfaces of the platform are generally open. Hence, a platform company offering a dominant platform design in its industry shapes a flexible platform-based ecosystem, which is open to a potentially unlimited pool of complementors. However, technological platforms may have distinctive forms and vary in the degree of their openness and the stakeholder types using it. Companies may also use technological platforms to provide selectively open access to the interfaces for their selected partners, thus forming a supply-chain platform. Furthermore, distributed company units may use internal technological platforms. They share the characteristics of a platform core and modular, extensible periphery, but the interfaces of the platform remain closed and are not accessible to external third party complementors (Gawer, 2014). The differing types of platform-related boundaries (varying between closed consortia or technical standard committees and open communities such as the ones for open-source software) and the different types of decision making (varying between heterarchical and hierarchical decision making) are also present in current platform manifestations (Gulati, Puranam, & Tushman, 2012). Besides, to understand how industry platforms work, platform architecture is another important topic of this sub-cluster. All technological platform types are understood as modular technological architectures, which can be expanded by modular complements. In digital markets, a platform is usually an extensible codebase of a software-based system that provides core functionality shared by apps that interoperate with the platform core, and the interfaces through which they interoperate (Tiwana & Konsynski, 2010). Thus, technological platforms align with the concept of software platforms, introduced by Tiwana and originate from information systems research.

Furthermore, from an architectural perspective, platforms are seen as a set of subsystems interacting through interfaces (Tiwana & Konsynski, 2010), sharing the characteristic of modularity. Modularity implies that the decoupling of platform modules from the platform core does not affect other modules (Tiwana & Konsynski, 2010). For the overall understanding of platforms, it is crucial to realize that digital products usually share the modular architecture as well (Yoo et al., 2010; Yoo et al., 2012). The same applies to the network externalities, which are recognized in the context of digital products. Influential research highlights how modularity is related to platform governance efforts, so evolutionary dynamics and specific rules are set in platform-based ecosystems to maintain control over the modular third-party contribution, simultaneously fostering the innovation (Tiwana & Konsynski, 2010). To improve control of complex modular platform systems and to manage the architecture and the governance, several representations, such as network graphs, design structure matrices, or layer maps are suggested (Baldwin & Woodard, 2009). By using mobile platform ecosystems as an exemplary industry, Basole and Karla (2011) demonstrate how useful visualizations of platform-based ecosystems are in order to understand their structure and dynamics. Furthermore, Reuver et al. (2018) points out the necessity to distinguish a large number of activities and variables in platform-based ecosystems and recommends more collaboration in the information systems community.

Also, Thomas et al. (2014) complete the holistic view on platforms and add the transactional logic. Through the provision of the digital infrastructure and the platform-related set of rules and interfaces, technological platforms also act as two-sided or multi-sided markets, building the second preliminary perspective. This perspective originates from economic research and conceptualizes platforms as a specific type of network, bringing together distinctive groups and facilitating the transactions between them, by coordinating demand and supply. Thomas et al. (2014) consider the transactional logic as a source of leverage, which is required to implement a platform in a market. Gawer (2014) also sees the intermediary role of platforms as integrated into the concept of technology platforms. However, the foundation of influential scientific publications in this area is included in the fourth cluster (#4 two-sided market) within the observed dataset.

Due to the openness of interfaces, the general understanding of technology platforms also requires an understanding of the ecosystem concept, which is highly influenced by Adner (2017) and Jacobides et al. (2018). Complementing the business ecosystem perspective, coined by Moore (1993), Adner (2017) conceptualizes ecosystems as a structure, in which a focal (platform) company is in the position to secure its role in a competitive ecosystem. The ecosystem strategy also describes how platform companies achieve this position (Adner, 2017). Jacobides et al. (2018) argue that platform-based ecosystems are a controllable process, sharing a similar perspective on ecosystems. Hence, platform companies need to understand what drives value and how to capture it based on rules. Empirical studies show that even relatively successful platform companies, such as Google, struggle with emerging sub ecosystems, which are less controlled by the platform provider. Thus, the understanding and the subsequent design of ecosystems are essential for the success of platforms (Jacobides et al., 2018).

Bridging the platform and the ecosystem concept, McIntyre and Srinivasan (2017) conduct a valuable literature study of platform-influenced networks from the perspective of the industrial organization economics. The authors provide a comprehensive overview of the platform-related contributions and highlight the specific industry settings of each study. The authors argue that different perspectives constrain the insights on how to design successful platforms. As an example, the authors mention the complexity of network dynamics, which exceeds the observation of network effects in specific industries. Furthermore, McIntyre and Srinivasan (2017) recommend focusing on the analysis of indirect network effects, which often remain unconsidered in prior studies (McIntyre & Srinivasan, 2017).

This sub-cluster also contains the most research agenda calls. Tiwana and Konsynski (2010) suggest the usage of four different theoretical lenses for further platform theory construction. Jacobides et al. and Reuver et al. also share the call for more platform theory research, adding various ideas for future research directions (Jacobides et al., 2018; Reuver et al., 2018). Yoo et al. (2010) develop a layered architecture of digitized products and consider them as being simultaneously a product and a platform. They use an Ipad as an exemplary digitized product to explain this dualism. The authors suggest a research agenda for the information systems research, which explicitly takes platforms into account as an essential component in a layered modular architecture of digitized products (Yoo et al., 2010). Therefore, the agenda covers the research areas of the coordination and the collaboration in platform-based ecosystems, referring to the concept of boundary resources (see sub-cluster 3). Furthermore, the agenda calls to seek explanations for companies' strategic choices related to platforms. Overall, it is worth mentioning that the period covered is eight years, and some research topics are still unaddressed. For instance, the concept of boundary resources is mentioned in the research agenda papers from 2010 and 2018 (Reuver et al., 2018; Yoo et al., 2010). Reuver et al. (2018) propose to use this concept for a better understanding of ecosystem dynamics, which is simultaneously disputed (McIntyre & Srinivasan, 2017). Furthermore, a deeper understanding of ecosystem governance is also highlighted in 2010 and 2018 (Jacobides et al., 2018; Yoo et al., 2010). We argue that despite the criticism of the specific industry observations, brought up by McIntyre and Srinivasan (2017), the specific context of each market is necessary for understanding governance mechanisms and their improvement. Three authors mention the need to understand the value creation (and the aligning value capture) mechanisms for every actor type in platform-based ecosystems (Jacobides et al., 2018; Thomas et al., 2014; Yoo et al., 2010). However, only Jacobides et al. (2018) propose a complementarity framework to kick off new research in this direction. Altogether, though some generic abstractions for governance and value creation were researched in the past, it remains largely unknown how the market-specific adaptions of these concepts should be designed (see section 4). Nevertheless, these research areas, such as governance mechanisms and influential factors on ecosystem dynamics, are partially addressed in the second and the third sub-clusters described below.

Sub-Cluster 2 is mainly based on the perception that company success depends not only on the companies themselves but also on their suppliers, distributors, outsourcing firms, makers of related products and services, technology providers, and additional partners, forming a business ecosystem (Iansiti & Levien, 2004). Jacobides, Knudsen, and Augier (2006) argue that firms can create an architectural advantage by fostering innovators in the ecosystem. They develop a framework that guides companies in extracting benefits from complementarity in different settings (Jacobides et al., 2006). However, Sheremata (2004) turns the tables and shows that ecosystem participants or competitors can also challenge existing ecosystems by introducing radical-incompatible innovation.

Hence, innovation is a strategy to oppose architectural advantages. Firms leveraging architectural advantages are hub firms of their innovation network. Therefore, the network recruitment process, as well as the network management and orchestration process of the hub firm plays an important strategic role in obtaining ecosystem stability and achieving desirable outcomes (Dhanaraj & Parkhe, 2006). Adner (2006) identifies restrictions in the management of innovation ecosystems and argues that firms should also mitigate risks imposed by dependencies from ecosystem partners, which might even lead to a revise of their strategy triggered by missing or delayed complementary products. To identify the underlying coherences between different complementary products, Adomavicius, Bockstedt, Gupta, and Kauffman (2007) introduce the concept of technology ecosystems, which divides complementary products of a technology ecosystem into different roles. Underling coherences and strategical implications can be derived whether a technological good is a component, a product/application, or support/infrastructure for other goods (Adomavicius et al., 2007). Gawer and Henderson (2007) confirm prior research on these topics with an extensive case study on Intel and elucidate how the organizational structure can have competitive consequences for the management of complementary markets. Additionally, Gawer and Henderson (2007, p. 6) define a platform owner "as a firm that owns a core element of the technological system that defines its forward evolution", thus introducing today's most common understanding of digital platforms. Afterward, Gawer and Cusumano (2008) introduce the concepts of coring and tipping as strategic options to become a platform leader. While coring offers technology and business actions to consider when creating a new platform, tipping offers guidance on how to gain momentum in platform markets to set an industry standard. Furthermore, it is crucial to recognize that standards by themselves are not platforms but an essential part of understanding which platform is likely to win the majority of a market (Cusumano, 2010).

Eisenmann et al. (2011) explore new strategies to enter platform markets. Their so-called "envelopment strategies" allow platform owners to combine their functionality with those of a new target market resulting in a multi-platform bundle that leverages shared user relationships. Furthermore, different envelopment strategies are presented based on whether platform pairs are complements, weak substitutes, or functionally unrelated (Eisenmann et al., 2011). Baldwin and Hippel (2011) support the idea of distributed innovation and point out that digitalization supports innovation by individual users and open, collaborative innovation, offering even more variability for the design of successful platform ecosystems. Explaining the growing importance of digital platforms, Yoo et al. (2012) introduce convergence and generativity as characteristics of pervasive digital innovation. Digital convergence combines previously separated technologies (e.g. internet, phone, and music), creates smart products by combining physical artifacts with digital technologies, and brings together formerly separated industries (e.g. internet and telecommunication). "Generativity means that digital technologies become inherently dynamic and malleable" (Yoo et al., 2012). This indicates that new capabilities can be added after the production of pervasive digital goods. Additionally, generativity leads to wakes of innovation (Boland, Lyytinen, & Yoo, 2007) and generates an enormous volume of digital traces as by-products, which can be used for further innovation and is commonly referred to as big data.

Ceccagnoli et al. (2012) analyze the impact of platform entry for independent software vendors (ISV) and find that the advantages of platform entry outweigh associated disadvantages. Additionally, their research shows strategic fields of action for ISVs, proving that greater intellectual property rights and downstream capabilities can be used to differentiate in the platform market and increases the business performance of ISVs Ceccagnoli et al. (2012). Gawer and Cusumano (2014) illustrate technological, strategic, and business challenges that platform leaders and their competitors face based on an intensive case study and develop effective practices for platform leadership. Additional strategical challenges for digital platforms from the perspective of multisided markets are presented by Hagiu (2014) who emphasizes that successful multisided platforms are the exception, pointing out that many platform companies fail due to the so-called chicken- and -egg problem, the resistance of key complementors and the high complexity of managing a platform ecosystem. Economic differences between multisided platforms, vertically integrated firms, and input suppliers and the respective strategic implications can be found in Hagiu and Wright (2015).

While most research on platform competition addresses inter-platform competition, Tiwana (2015) examines intra-platform competition and reveals the importance of architectural control imposed by the platform owner, to increase compatibility and market performance of complementors. Boudreau and Jeppesen (2015) analyze the behavior of unpaid crowd complementors and find that platforms might develop positive indirect network effects while developing sufficient negative direct network effects leading to net zero effects and what they call the network effect mirage.

Van Alstyne et al. (2016) summarize the insights above and explain differences between traditional pipeline firms and platforms as well as the strategic implications. A different focus can be found in Eloranta and Turunen (2016), who convey the insights of platform research to service networks in manufacturing to leverage network-related complexity in their operations. The last paper in this cluster shows that platforms often need new strategies for internationalization since new markets and regions have no connection to the existing network effects (Brouthers et al., 2016).

An overall evaluation of sub-cluster 2 shows the linking character of the strategic perspective on platforms. Publications in this sub-cluster apply the definitions and principles of the rather generic sub-cluster 1 to specific industry settings. These settings enable the authors to define desirable outcomes of platform ecosystems. Therefore strategies, how to achieve these goals can be developed on a conceptual level. However, these strategies set the foundation for empirical research in sub-cluster 3.

Drawing on the perception that the success of platform-based ecosystems is a result of a design process, sub-cluster 3 focuses on relevant determinants and strategies to improve the platform or its ecosystem. This cluster consists of 31 scientific papers and two management books. The main criterion for this cluster is, whether the paper offers specific strategies or frameworks to improve the technical or non-technical design of a platform or its aligning ecosystem.

The first step of the review procedure relies on the methodical approach defined by Webster and Watson (2002). Consistent with the suggestion of Webster and Watson (2002), we apply a theoretical framework to determine which platform characteristics or strategies, the authors consider important in the analyzed scientific articles. We identify the following seven research dimensions based upon the content analysis and inductive coding: technical platform architecture, compatibility and standardization, complementarity, participation architecture, boundary resources, property rights, and pricing. In the following, we analyze the articles in chronological order to get an overview of the research path of sub-cluster 2. The following matrix sums up the identified research dimensions and the publications, matching at least one of the dimensions:

*Table 4: Concept matrix for the development of platform ecosystems*

| Authors / Concepts | Platform Architecture | Compatibility (Standardization) | Complementarity | Participation Architecture | Boundary Resources | Property rights | Pricing |
|---|---|---|---|---|---|---|---|
| Nair, Chintagunta, and Dubé (2004) | x | | | x | | | |
| Venkatraman (2004) | | | | x | | | |
| Au and Kauffman (2008) | | | | x | | | |
| West and O'mahony (2008) | | | | x | | x | |
| Gallagher and West (2009) | | x | | | | | |
| Boudreau and Hagiu (2009) | | | | x | | | x |
| Adner and Kapoor (2010) | | | x | | | | |
| Boudreau (2010) | | | | x | | | |
| Xu, Venkatesh, Tam, and Hong (2010) | | | x | | | | |
| Shy (2011) | | x | | | | | |
| Nambisan and Sawhney (2011) | | x | x | x | | | |
| Zhu and Iansiti (2012) | x | | | | | | x |
| Boudreau (2012) | | | x | x | | | |
| Carare (2012) | | | | | | | x |
| Cennamo and Santalo (2013) | | | x | x | | | |
| Ghazawneh and Henfridsson (2013) | | | | | x | | |
| Huang, Ceccagnoli, Forman, and Wu (2013) | | | x | | | x | |
| Henfridsson and Bygstad (2013) | | x | x | | | | |
| Cenamor et al. (2013) | | | x | | | | |
| Tiwana (2014) | x | x | x | x | | | |
| Anderson, Parker, and Tan (2014) | x | | | | | | x |
| Wareham, Fox, and Cano Giner (2014) | | x | x | | | | |
| Lee and Raghu (2014) | | | | | | | x |
| Ghose and Han (2014) | | | | | | | x |
| Kang and Downing (2015) | x | | | | | | |
| Benlian, Hilkert, and Hess (2015) | x | x | x | x | | | |
| Ondrus, Gannamaneni, and Lyytinen (2015) | | | | x | | | |
| Eaton, Elaluf-Calderwood, Sørensen, and Yoo (2015) | | | | | x | | |
| Parker et al. (2017) | | | x | x | | | |
| Parker and van Alstyne (2018) | | | | x | | x | |
| Kapoor and Agarwal (2017) | | | x | x | | | |
| Zhu and Liu (2018) | | | x | x | | | |
| Kazan et al. (2018) | | | | x | | | |
| **Cumulated Topics** | **6** | **7** | **13** | **17** | **2** | **3** | **6** |

Examined chronologically, earlier papers in this cluster utilize the characteristics of the technology platform concept, without citing Gawer and Cusumano (2002). The platform concept is used to describe exemplary networked markets, such as personal digital assistants (PDA) and game consoles, to explore network effects (Nair et al., 2004; Venkatraman, 2004). In their empirical research, Nair et al. (2004) develop an econometric framework and research the dependence of PDA hardware on compatible software, taking into account the technical architecture of PDA devices. Venkatraman (2004) identifies four factors of influence for complementors to join a platform, namely the density overlap, the embeddedness of a platform, as well as the dominance and the newness of a platform. By studying open source communities, West and O'mahony (2008) coin the term of participation architecture and develop a framework, which consists of three design parameters: production, governance, and intellectual property. Similar to a sponsored open source community, platform companies may utilize these design parameters to establish a platform-based ecosystem (West & O'mahony, 2008). This paper also sheds light on property rights in platform strategy research. Furthermore, Au and Kauffman (2008) address differing stakeholder characteristics in platform-based ecosystems. By developing a stakeholder model, Au and Kauffman (2008) describe several stakeholder-related economic issues and theories (e.g. switching costs, network externalities) and illustrate how they differ for each stakeholder type. Furthermore, the cluster considers the changing nature of competition and the role of standards and standardization in platform-driven markets. Gallagher and West (2009) apply the positive loop model and analyze its impact on the adoption of standards.

The most influential work in this sub-cluster was conducted by Adner and Kapoor (2010). Their research article creates awareness for external effects influencing the ecosystem and possible consequences for the platform provider, issued by technological interdependence. Building up the context of internal and external innovation challenges and bundling them into a framework, Adner & Kapoor (2010) extend research on complementarity. They identify components and complements as sources of external challenges, as both are provided either by suppliers or by complementors. If the challenges in the ecosystem are related to components, it may improve the competitive advantage of the platform provider. However, if the challenges are related to the complements, they may reduce the competitive advantage of the platform provider. Hence, platform companies should consider their position in the ecosystem and the technological interdependence to analyze technological integration challenges of external components and the challenges of critical complements, which may result in value reduction of the complete platform. In his 2010 paper, Boudreau (2010) studies the innovation impact of opening a platform and enhances prior knowledge about the effects of the participation architecture, dividing it into two stages: granting access through opening the platform and giving up the control. This implies sharing intellectual property and collaborating with complementors on shared projects. The empirical study shows that opening the platform has a more substantial positive effect on innovation by complementors, than giving up platform control, which has only a small positive effect. In his latest paper within the sub-cluster, Boudreau (2012), uses a similar data set of handheld devices to study the variety of software in platform-based ecosystems and proofs, that the innovation rates may vary and distinctive and specialized complementors should be favored. Furthermore, the paper discovers the link between the willingness of complementors to innovate and the differentiation level of complementary software. Thus, the evidence shows that the incentives to innovate are low for complementors with a higher number of substitutes. Consequently, the amount of complementors changes the nature and sources of innovation but does not necessarily increase the innovation of the ecosystem.

Research into the migration behavior of end customers on a platform and the resulting platform adaption is rare. The only identified model classifies technology perceptions, external influences, and complementarities as influencing categories for end customers to migrate in platform-networked markets. Furthermore, this paper conceptualizes the complementarities between hardware and software platforms (Xu et al., 2010). Additionally, a paper written by Shy (2011) gives a comprehensive list of important market specifics, influenced by network effects. Orchestration of innovation output in platform-networked markets is recognized as an interplay between elements of innovation design and network design and combined as an orchestration model, consisting of several strategies for platform companies (Nambisan & Sawhney, 2011). Zhu and Iansiti (2012) empirically prove how technical quality advantage of a platform in combination with indirect network effects and discounts for customers, issued by the platform-provider, positively impact ecosystem development and the aligning complementor success and may build a market-entry strategy for starting platforms. Capturing the end-customer perspective, Carare (2012) analyzes pricing structures and optimal pricing levels in mobile application stores and reveals the relationship between the willingness of end customers to pay and the position in the app store. This paper is one of the few within the cluster to tackle pricing in platform ecosystems. The fifth most influential paper by (Cennamo & Santalo, 2013) draws on the winner takes it all assumption and analyzes how platform performance suffers, in a setting where several platforms coexist in one market, pursuing the same strategy, and demand exclusive agreements with complementors, prohibiting them from supporting more than one platform. The authors also argue that a distinctive platform strategy is more fruitful in that case than restricting multi-homing.

Ghazawneh and Henfridsson (2013) develop a model for the previously rather generic understanding of the participation architecture and apply the boundary object theory on software platforms, which results in the boundary resource model. This model can be used by platform companies to govern the ecosystem, either fostering or securing the innovation capabilities of the platform (Ghazawneh & Henfridsson, 2013). Another study of the video game console market, conducted by Anderson et al. (2014), reveals how platform companies may overinvest in the performance of the platform core and how important it is for the platform company to get a detailed understanding of the needs of customers and complementors to develop a leading platform. Huang et al. (2013) shift the attention of complementors and platform companies to intellectual property rights (IPR), arguing that such formal mechanisms are significant to prevent complementor expropriation. Thus, the establishment of an appropriate level of IPR is an important factor for potential complementors to protect themselves. Therefore, the understanding of IPR related effects by platform companies might improve complementor engagement as well. Cenamor et al. (2013) analyze platform adaption in the video console market and reveal the importance of complementary products, as they drive dynamics in platform-based markets. Referring to the generativity effect of platforms, conceptualized as a digital infrastructure, prior research recognized adoption, innovation, and scaling as strategies to foster generativity Henfridsson and Bygstad (2013). The management book "Platform Ecosystems: Aligning Architecture, Governance, and Strategy" by Tiwana (2014) deserves special attention as it contains a comprehensive view on platform-related concepts (e.g. platform architecture, governance, and dynamics) and sums up several platform strategies how to evolve the platform and succeed in the orchestration of external innovation.

Wareham et al. (2014) present a list of specific mechanisms and strategies to manage complementary output. Exemplary mechanisms to foster the output include certifications, training, code sharing, technical, and financial support. Orchestration strategies for heterogeneous complementors and their capabilities enhance the research dimension of the participation architecture. Lee and Raghu (2014) prove how the probability of conquering the app store ranking increases, if complementors offer a wide variety of applications across multiple categories of the application store, thus providing platform-related strategic considerations for complementors. Another pricing study focuses on the impact of discounts and in-app advertisements on the sales performances of mobile applications (Ghose & Han, 2014). Kang and Downing (2015) highlight several relevant influential factors regarding the successful entrance in platform-organized markets and argue that the keystone advantage (mentioned in sub-cluster 2) is decisive for a newly launched platform and helps to overcome the lacking user base in a nascent ecosystem.

Two papers from 2015 focus on platform openness to improve platform-based ecosystems. Research work with a focus on openness provides a model of perceived platform openness (PPO) to improve complementor engagement (Benlian et al., 2015). The PPO model is based upon some dedicated factors for perceived openness, such as the technical documentation, the support, or the technical performance of the platform. Prior research additionally provides three different types of platform openness and explores the effect of the resulting openness configurations, based upon the exemplary market of mobile payment (Ondrus et al., 2015). The identified openness configurations extend over three levels of openness, such as provider, technology, and user levels. The paper proposes a decision model, which helps platform providers to decide whether opening the platform is useful or not. Both papers address fundamental problems of the participation architecture and develop approaches for its optimization. In addition, Eaton et al. (2015) draw on prior research of boundary resources and prove empirically, how Apple managed to change and adjust its boundary resources (actions of distributed tuning), uncovering the evolutionary perspective on this participation architecture dimension. Furthermore, the paper analyzes the power distribution in platform-based ecosystems and finds evidence that complementors are in the position to exercise power and influence the adjusting process of boundary resources.

The examination of recently published papers in this cluster exposes interesting rules on the improvement of platform ecosystems. In their paper, Parker et al. (2017) build a model for optimization of third-party innovation in platform ecosystems under consideration of competition and developer risk, contributing to the research dimensions of complementarity and the participation architecture. In a follow-up paper, van Alstyne and Parker focus on managerial decisions about the effects of closing the platform and charging for platform access. Furthermore, this paper researches the most favorable duration of property rights for complementors and analyzes the ecosystem-wide effects of the expiration of intellectual property rights (e.g. code spillover) (Parker & van Alstyne, 2018). Kapoor and Agarwal (2017) discover ecosystem complexity, indicated by the structural (i.e. technological interdependencies) and evolutionary (i.e. technology shift caused by the platform provider) features. The study shows how evolutionary features positively correlate with the superior performance of complementors. Furthermore, their paper shows that the transitions of the platform technology negatively affects complementor performance. Zhu and Liu (2018) explore the case of Amazon and depict a tension between the acquisition of complements to ensure platforms' long-term growth and short-term incentives from third-party activities. These insights improve the strategical decisions of both platform providers and complementors (Zhu & Liu, 2018).

In the latest publication, Kazan et al. (2018) use value creation and value delivery architectures as dimensions of a taxonomy to classify platform providers and extract three possible platform competition strategies (Kazan et al., 2018). In this descriptive study of the mobile payment domain, the authors characterize monopolistic and co-opetitive strategies, which rely on rather traditionally established architectures and a transformation strategy, built upon product and process innovations.

The cumulative values (see Table 4) indicate that the participation architecture and the study of complementarities are the core dimensions within the strategy-focused body of platform literature. These two dimensions are the most discussed. The discussion of distinctive platform architecture strategies (also defined as platform- or ecosystem governance by Tiwana) was a dominant topic of the platform research community between 2004 and 2008 and since 2015. Between these two periods, the researched dimensions are varying. However, one can notice, specific governance strategies or participation architecture propositions are relatively rare. The earliest contributions about the participation architecture and the recently published papers on this topic are mostly conceptual or propose theoretical models. Empirical studies mostly apply case studies and secondary data to define the determinants and the effects of the participation architecture. Ghazawneh and Henfridsson (2013) build the only concrete concept to manage the heterogeneity in platform ecosystems and to regulate the access to the platform functionalities. That is why we recognize boundary resources as an independent research dimension. Apart from this concept, Wareham et al. (2014) propose specific mechanisms to foster the variety of external complements, and Benlian et al. (2015) develop clear guidance for platform design with their perceived platform openness model. The current focus of the research on the participation architecture suggests a keen interest in investigating determinants of effective platform strategies.

Sub-cluster 4 is complementary (excluded in table 4 due to lower relevance) and consists of 29 papers and five management books, addressing various additional research streams that more or less relate to research on digital platforms. According to the conducted content analysis, most cited literature offers the innovation management focus and explores how digital platforms and platform-based ecosystems affect it. Due to the collective efforts as well as the external pool of resources, these perspectives have a certain overlap with the perspectives of platform providers and complementary third-parties. This is partly caused by the inability of single ecosystem participants to understand the overall scope of innovation in multi-layered digital products, fostered by flexible platform ecosystems (Nambisan, Lyytinen, Majchrzak, & Song, 2017). Nevertheless, the innovation management builds the scope of this research perspective and platforms only define the setting, without being considered a research object.

Similarly, entrepreneurship ecosystems build an additional research object (Clarysse, Wright, Bruneel, & Mahajan, 2014). One current research question is how digital platforms can support entrepreneurship. Hsieh and Wu (2019) offer a framework to address specific innovation challenges with suitable platform types to support product/service development and commercialization with platform strategies. Furthermore, the cluster contains essential work for the digitization of companies (Bharadwaj, 2013), which is not focused on platforms and defined too broadly. In summary, we argue that the literature contained in this cluster is not necessarily crucial for the understanding of how digital platforms work and how platform-based ecosystems can be established from the platform-provider perspective or how complementors and customers as platform users should behave in platform-based ecosystems to sustain and to succeed in value capturing.

## 4. Discussion and objectives for future research

In this section, we discuss possible future research topics, based upon the qualitative content analysis of the platform ecosystem cluster. The examined body of literature contains three distinctive research agendas, whereby two of them are dated to the year 2018. Reuver et al. (2018) plead for new theoretical contributions related to platform development, platform governance, a transformation of industries through platforms, and new design-related theories. Compared with prior research in this cluster, future research needs precise units of analysis and research scoping, utilizing rarely used methodologies, such as longitudinal studies, comparisons between several platform case studies, visualization techniques, and data-driven approaches (Reuver et al., 2018). These research implications extend the research topics proposed by Yoo et al. (2010). We came across several studies, researching platform ecosystem improvement, which used secondary data to conduct longitudinal studies (Adner & Kapoor, 2010; Boudreau, 2010; Cenamor et al., 2013; Eaton et al., 2015). Compared to the total number of articles contained in the cluster, this method is nevertheless only used relatively rarely. Graphical networking is also only used in very few studies (Basole & Karla, 2011; Clarysse et al., 2014). We did not discover the usage of any data science methods or simulations. It is also worth mentioning that the case study research design, elaborated by Yin (2014), is commonly cited in the whole cluster. Hence, case studies are the most commonly used research method to empirically test the hypothesis and conceptual models in the platform context. This insight leads us to the proposition to extend the variety of applied research methods and to carry out more business simulations (e.g. system dynamics or multi-agent simulations), data science methods to mine electronic documents (Bowen, 2009) or prototyping studies to address the research agendas. Jacobides et al. (2018) do not propose any specific research methods, but two contextual overlaps are evident. Firstly, the coordination- and governance-related implications indicate that this research avenue should be continued building on the developed and validated concepts (Boudreau, 2010; Eaton et al., 2015; Ghazawneh & Henfridsson, 2013; Wareham et al., 2014). Value capture as a research topic can be combined with the deeper integration of cluster #10. The core business model characteristics such as value proposition and value capture match with the value creation research direction and might improve the development of platform ecosystems as well. The integration of the business model lens can foster the transparency of revenue stream sharing and might help to integrate the platform design characteristics and policies, which affect the value capture mechanisms and can negatively influence the attractiveness of a platform against the background of competitive dynamics (Jacobides et al., 2018). Because empirical research on digital platforms has grown enormously over the last years, we sense an opportunity to enrich existing platform strategies with empirical evidence. Detailed findings on technical aspects of platforms, as well as current insights on the relationship between the different user groups on platforms, may lead to comprehensive strategies. These strategies might not address desirable generic outcomes but rather specific in and output dimensions of platform ecosystems.

Similar to the PPO model, a perceived platform quality model can also be a future research object. The examined studies do not consider the maturity of the market in this context. Nascent platforms in business-to-business markets (e.g. industrial Internet of Things, cloud platforms) or domains, where hardware is not proprietarily bound to a specific software platform require additional attention, due to changing factors (e.g. a shift in ecosystem complexity and openness levels) of influence to evolve a platform-based ecosystem. Drawing on the platform architecture control, researched by Tiwana (2015), we need a deeper understanding of how to keep the balance between the stimulation and the control of third-party innovation from the platform provider perspective. Although Ghazawneh and Henfridsson (2013) conceptualized BR for these two opposing effects, specific design guidelines are missing to manage this tradeoff. This research implication is consistent with the research call for platform design from Reuver et al. (2018) and could benefit from the application of data science methods, such as repository mining, to capture the preference of platform users. Furthermore, an instantiation of the concept for different usage scenarios for platforms from clusters #2, #6, #7, #8, #15 would be delightful for the development of application-specific platform strategies. Also, boundary resources represent a possibility to open a platform (Karhu, Gustafsson, & Lyytinen, 2018). In our analysis of the platform ecosystem cluster, we did not find any managerial recommendations about the required circumstances for a company to successfully open a closed platform ecosystem, offering an additional field for future research.

Furthermore, research insights into the effects of multi-homing are not clear. It is still unclear in which situations multi-homing should be prevented or promoted either by the platform provider or the complementor. For example, multi-homing shows the complex contexture between strategical and technical aspects in platform ecosystems. Generativity and complementarity define specific technical characteristics of the platform, predefining strategic decisions on multi-homing. Current research misses guidance on how business strategies can be adequately addressed in the technical development of platforms. Since complementors and end customers are considered valuable resources for platform companies, more studies might also research the connection between envelopment (Eisenmann et al., 2011) and the architectural design of platforms. A deeper understanding of specific characteristics of the technical platform architecture or platform control mechanisms might prevent envelopment on an architectural level in the future. Embedded case studies could be beneficial for the exploration of such settings.

Surprisingly, IPR is a research object in only four papers. Possible future directions of research may include specific design criteria for IPR and develop models, supporting platform companies to create optimal incentives for complementors to join. It can be assumed that the emergence of an independent literature cluster with a research focus on intellectual property rights supports the consideration of IPR as an essential research object in management and business strategy.

Only six studies focus on platform-related pricing issues, three of which capture the complementor perspective and formulate recommendations for the pricing of apps. Against this background, a platform provider may distinguish pricing strategies and pricing levels depending on the level of stakeholder participation (Wareham et al., 2014). Because pricing is in the focus of network cluster #4 "multisided markets", researchers in cluster #2, as well as practitioners, can rely on these findings and use the insights to develop appropriate pricing strategies in platform ecosystems under consideration of market dynamics. Unfortunately, our analysis indicates that many of the achievements in the economic research domain have not yet devolved into distinct recommendations for the governance of specific platforms. We argue that the knowledge about the pursued strategies of different platforms, which compete in a particular market, also represents essential knowledge for complementors. The application of the platform user perspective as an additional lens to the systematical literature analysis of the platform ecosystem cluster reveals that the five most influential articles in this sub-cluster primarily represent the perspective of platform companies. In sum, only eight papers provide significant contributions for platform users, which is less than 7,5% of the whole cluster. Hence, the specific needs and requirements of platform users are insufficiently addressed and offer another critical direction for future research.

Furthermore, we find a need to study the inner life of platform ecosystems. Our research shows general insights about structures and the orchestration of ecosystems. However, there is a lack of research describing the inner life of platform ecosystems. More specifically, there is a lack of studies explaining the interplay between different platform users on a user level. A platform ecosystem embodies an extensive network of interconnected users, which is rarely examined regarding the dynamics between different users and different user groups. Analyzing network effects on a micro-level (e.g. user-level, group-level) enhances understanding of the dynamics of platform ecosystems. Lacking insights about micro-level effects on platforms restrict the development of useful governance and coordination mechanisms.

Lastly, we identified an additional research stream, which takes into account the nascent entrepreneurship ecosystems (Clarysse et al., 2014; Thomas et al., 2014). This research stream considers platform-based ecosystems as a setting and may develop in the future. Similarly, research clusters with a context-specific platform usage (#0, #6, #7, #8, #15) might require a more in-depth look at how platforms can improve phenomena such as entrepreneurship, social media or crowdfunding. Furthermore, it would enhance the understanding of how platforms influence competition and how they transform these phenomena. Such a research direction aligns with Reuver et al. (2018) requesting to conduct more embedded case studies to understand the specific role of platform ecosystems in each context.

## 5. Conclusion and limitations

This study represents an attempt to systematize the existing body of research on the concept of platforms as defined by Gawer and Henderson (2007) from the perspective of management research and business economics research to improve understanding of different convoluted concepts shaping the body of knowledge on digital platforms. We discover 14 platform-related research fields. These encompass fundamental research fields, which have formed the business economics perspective on digital platforms as well as new and developing research fields, which mainly address specific technical use cases of platforms. Although these clusters resonate with previous reviews, these clusters are based on co-citation patterns, which are more objective than the subjective evaluations used to identify similar structures in previous reviews. By taking a detailed look at the "platform ecosystem" cluster, our quantitative bibliometric analysis finds 107 papers closely connected to this research stream. The analysis of these papers identifies three distinct objectives in the "platform ecosystem" research cluster. First, research intended to define or unify the understanding of digital platforms. Second, papers aiming to improve companies via platform strategies. Third, studies that analyze specific characteristics of platforms to improve these. Development of the research objectives helps to reduce the ambiguity surrounding platforms and related concepts (e.g. see sub-cluster 4) in the scholarly discourse.

In terms of theoretical implications, this study contributes to the current body of literature on platforms and platform ecosystems, by analyzing the evolution of this field of research, trends, and emerging topics that are under-represented and require additional investigation. The study demonstrates the use of a proven bibliometric co-citation methodology to contribute to the body of knowledge on digital platforms and creates a boundary object for the business economics research community (Raghuram et al., 2010), highlighting main research objects, their development, and future research topics. Based on our extensive literature study, we invite other researchers to build on previous findings from associated research fields to develop in-depth findings across different disciplines in order to minimize conceptual ambiguity and refine business economics research on digital platforms.

Our study is subject to several limitations. First, the issue of all citations being treated equally when, in fact, they are not is a significant source of criticism. The rationale for citing a study could vary considerably, ranging from a reference that supports the work to criticism or the use of the same methodology (Baumgartner & Pieters, 2003). Another limitation of our study is that co-citation analysis favors older publications because they have been cited more often than current papers. This is a minor problem for the identification of our research clusters because they built up over a long period. Additionally, this leads to the disadvantage that recent publications could be excluded from the analysis due to their lack of co-citations.

Second, we have excluded most of IS research[2] from our data set due to the enormous body of IS literature related to any platform. Therefore, we have only found IS research that has already been connected to the business economics perspective on digital platforms, which indicates that other fruitful concepts, which could improve the understanding of platform ecosystems, might still be hiding in IS research. Therefore, this study provides IS researchers with successful concepts that have already been applied to platform research. This should consequently motivate IS researchers to enhance business research on digital platforms with new theories from IS research. Software ecosystems (SECO) represent one such exemplary concept. The advancement of the concept of business ecosystems (Moore, 1993) was defined by Jansen, Finkelstein, and Brinkkemper (2009). Specifics of the software industry supplement the non-software specific definition of ecosystems in this concept. The measurement of the potential of SECO goes back to Iansiti and Levien (2004) (see section 3.2), while Cusumano was involved in the development of the SECO definition (Jansen & Cusumano, 2013). The more surprising it is that the SECO concept is not included in the examined body of literature. This concept may help to link the two research areas. Hence, despite the mentioned limitations, we hope that this study provides valuable results supporting further research on digital platforms.

---

[2] Additionally, conference proceedings (e.g. International (A) and European (B) conferences on Information Systems) are not considered in our analysis.